\documentclass[pre,twocolumn,showpacs,preprintnumbers,amsmath,amssymb]{revtex4-1}

\usepackage{graphicx}
\usepackage{dcolumn}
\usepackage{bm}

\begin{document}

\newcommand{\bea}{\begin{eqnarray}}
\newcommand{\eea}{  \end{eqnarray}}
\newcommand{\bit}{\begin{itemize}}
\newcommand{\eit}{  \end{itemize}}

\newcommand{\be}{\begin{equation}}
\newcommand{\ee}{\end{equation}}
\newcommand{\ra}{\rangle}
\newcommand{\la}{\langle}
\newcommand{\U}{\widetilde{U}}


\def\bra#1{{\langle#1|}}
\def\ket#1{{|#1\rangle}}
\def\bracket#1#2{{\langle#1|#2\rangle}}
\def\inner#1#2{{\langle#1|#2\rangle}}
\def\expect#1{{\langle#1\rangle}}
\def\e{{\rm e}}
\def\proj{{\hat{\cal P}}}
\def\tr{{\rm Tr}}
\def\H{{\hat H}}
\def\Hdag{{\hat H}^\dagger}
\def\Lop{{\cal L}}
\def\Ehat{{\hat E}}
\def\Edag{{\hat E}^\dagger}
\def\Shat{\hat{S}}
\def\Sdag{{\hat S}^\dagger}
\def\Ahat{{\hat A}}
\def\Adag{{\hat A}^\dagger}
\def\U{{\hat U}}
\def\Udag{{\hat U}^\dagger}
\def\Zhat{{\hat Z}}
\def\Phat{{\hat P}}
\def\Op{{\hat O}}
\def\id{{\hat I}}
\def\x{{\hat x}}
\def\P{{\hat P}}
\def\Px{\proj_x}
\def\Pr{\proj_{R}}
\def\Pl{\proj_{L}}


\title{Classical to quantum correspondence in dissipative directed transport}

\author{Gabriel G. Carlo}
\affiliation{Departamento de F\'\i sica, CNEA, Libertador 8250,
(C1429BNP) Buenos Aires, Argentina}
\author{Alejandro M. F. Rivas}
\affiliation{Departamento de F\'\i sica, CNEA, Libertador 8250,
(C1429BNP) Buenos Aires, Argentina}
\author{Mar\'\i a E. Spina}
\affiliation{Departamento de F\'\i sica, CNEA, Libertador 8250,
(C1429BNP) Buenos Aires, Argentina}

\email{carlo@tandar.cnea.gov.ar,rivas@tandar.cnea.gov.ar,spina@tandar.cnea.gov.ar}

\date{\today}

\pacs{05.45.Mt, 03.65.Yz, 05.60.Gg}

\begin{abstract}

We compare the quantum and classical properties of the (Quantum) Isoperiodic Stable Structures -- (Q)ISSs --
which organize the parameter space of a paradigmatic dissipative ratchet model, i.e. the dissipative
modified kicked rotator. We study the spectral behavior of the corresponding classical Perron-Frobenius
operators with thermal noise and the quantum superoperators without it for small $\hbar_{\rm eff}$ values.
We find a remarkable similarity between the classical and quantum spectra. This finding significantly extends
previous results -- obtained for the mean currents and asymptotic distributions only -- and on the other
hand unveils a classical to quantum correspondence mechanism where the classical noise is qualitatively
different from the quantum one. This is crucial not only for simple attractors but also for chaotic ones,
where just analyzing the asymptotic distribution reveals insufficient. Moreover, we provide with a detailed
characterization of relevant eigenvectors by means of the corresponding Weyl-Wigner distributions, in order
to better identify similarities and differences. Finally, this model being generic,
it allows us to conjecture that this classical to quantum correspondence mechanism is a universal
feature of dissipative systems.

\end{abstract}

\maketitle

\section{Introduction}
\label{sec1}

Research in directed transport has been steadily evolving since the main idea re-emerged in the 60's
\cite{Feynman} to consolidate into a well established field nowadays \cite{Reimann,Kohler,Denisov}.
Out of equilibrium spatiotemporal periodic systems are at the heart of these investigations,
where the second law of thermodynamics no longer applies. Many disciplines have found the concept
of ratchet transport very useful, they include such different areas as biology \cite{biology} on
one end and nanotechnology \cite{nanodevices} on the other. In chemistry, for example isomerization
reactions have been recently characterized by directed transport properties \cite{chemistry}.
Cold atoms \cite{CAexp,AOKR} and also Bose-Einstein condensates have been transported, these latter
by means of quantum ratchet accelerators \cite{BECratchets}, where
the current has no classical counterpart \cite{purelyQR} and
the energy grows ballistically \cite{QR2,coherentControl}. This short list is incomplete and
many other theoretical and experimental areas have been omitted here.

Breaking all spatiotemporal symmetries leading to momentum inversion \cite{origin} allows
a net current. In particular, we are interested in deterministic ratchets with dissipation
which have been historically associated with a classical asymmetric chaotic attractor \cite{Mateos}.
Dissipative quantum ratchets, interesting for cold atoms experiments
have been introduced in \cite{qdisratchets}. In recent works, the parameter space of the classical
counterpart of this system has been studied in detail \cite{Celestino}. These new results have pointed
out that not only chaotic attractors, but also several families of isoperiodic stable
structures (ISSs, which are Lyapunov stable islands of different periods grouped into
structures in parameter space, sometimes called ``shrimps'' due to their shape) have a
fundamental role in understanding the current behavior. Moreover,
they have been identified as a means to obtain optimal ratchet transport which is temperature
resistant \cite{Manchein}. It is worth noticing that these ISSs are a common feature,
found in generic dissipative dynamical systems. In fact, some of their properties have recently
been studied in the dissipative kicked rotator model \cite{Robnik}.

Their quantum manifestations, the so-called QISSs have been
investigated in \cite{Carlo}. There it has been found that the
QISSs look like the quantum chaotic attractors at their vicinity
in parameter space (these corresponding to values of the
parameters where the classical counterparts are already chaotic),
with the exception of comparatively few cases. Recently,
a complete parameter space picture has been obtained
for the quantum system \cite{Ermann}. It was also shown in \cite{Carlo} that a
thermal coarse-graining of the classical dynamical equations
(i.e., adding thermal noise of the order of $\hbar_{\rm eff}$) is
sufficient to obtain a good approximation for the QISSs. In this
paper we carry out a detailed spectral study of both the
Perron-Frobenius operators associated with the classical evolution
with thermal noise and the quantum superoperators without it,
finding a great similarity between them. A
known result in the decoherence literature \cite{nonnen} states
that diffusive noise makes the quantum mechanical spectrum
converge to that of the coarse grained Perron-Frobenius, the
diffusive process being applied to both the classical and the
quantum cases. Here, we find that for dissipative systems an
effective way to reach the classical to quantum correspondence
consists in applying a thermal (diffusive-like) noise solely at
the classical level. We underline that this is not meant to be an
example of the emergence of classicality \cite{Zurek}. It is
important to say that in this work all the possible representative
asymptotic cases have been considered, i.e. those in which the
classical limiting set is either simple (in its main forms) or
chaotic. Nevertheless a systematic study is left for the future.
Finally, we study the similarities
and differences of relevant eigenvectors with the help of
Weyl-Wigner distributions in phase space. Thanks to this we are
able to identify the limits of this correspondence mechanism.

This paper is organized as follows: In Sec. \ref{sec2} we present the system under
study, i.e. a modified kicked rotator with dissipation and the methods we use to
investigate it from the classical to quantum correspondence perspective. We
introduce the Perron-Frobenius operator and the quantum superoperator. In Sec. \ref{sec3}
we analyze their spectra, showing the classical to quantum correspondence details.
In Sec. \ref{sec4} we look at the behavior at the phase space level
by means of the Weyl-Wigner distributions of the eigenvectors. Finally,
in Sec. \ref{sec5} we present our conclusions.

\section{Model and calculation methods}
\label{sec2}

We investigate the paradigmatic dissipative ratchet system given by the map \cite{qdisratchets,Manchein}
\begin{equation}
\left\{
\begin{array}{l}
\overline{n}=\gamma n +
k[\sin(x)+a\sin(2x+\phi)],
\\
\overline{x}=x+ \tau \overline{n},
\end{array}
\right.
\label{dissmap}
\end{equation}
where $n$ is the momentum variable conjugated to $x$, $\tau$ is the period of the map
and $\gamma$ is the dissipation parameter.
This represents a particle moving in one dimension
[$x\in(-\infty,+\infty)$] under the influence of a periodic kicked asymmetric potential:
\begin{equation}
V(x,t)=k\left[\cos(x)+\frac{a}{2}\cos(2x+\phi)\right]
\sum_{m=-\infty}^{+\infty}\delta(t-m \tau),
\end{equation}
where $\tau$ is the kicking period, also having a dissipation
given by $0\le \gamma \le 1$. When $\gamma=0$ the particle is in
the overdamped regime and when $\gamma=1$ the evolution is
conservative. Breaking the spatial ($a \neq 0$ and $\phi \neq m
\pi$) and temporal ($\gamma \neq 1$) symmetries allows the net
current generation. As is customary in this model the classical
dynamics can be made dependent on the parameter $K=k \tau$ by
means of introducing the rescaled momentum $p=\tau n$. In order to
consider a thermal noise of the order of $\hbar_{\rm eff}$,
the effective Planck constant which will be introduced in the
quantum version, we simply change $\overline{n}' \rightarrow
\overline{n}$ in Eq. \ref{dissmap}, where
$\overline{n}'=\overline{n} + \xi$. We can associate the noise
variable $\xi$ with a temperature $T$ by means of the relation $
<\xi^2> =2 (1-\gamma) k_B T$, where $k_B$ is the Boltzmann
constant (which we take equal to 1). The factor $(1-\gamma)$ has
been taken to avoid having noise in the conservative limit,
however many other choices are possible. Finally we have that $T
\simeq \hbar_{\rm eff}/[2 (1-\gamma)]$.

Quantizing this model (without thermal noise) is straightforward:
$x\to \hat{x}$, $n\to \hat{n}=-i (d/dx)$ ($\hbar=1$).
Since $[\hat{x},\hat{p}]=i \tau$, the effective Planck constant
is $\hbar_{\rm eff}=\tau$. The classical limit corresponds to
$\hbar_{\rm eff}\to 0$, while $K=\hbar_{\rm eff} k$ remains constant.
Dissipation at the quantum level is introduced by means of the
master equation \cite{Lindblad} for the density operator $\hat{\rho}$ of the
system
\begin{equation}
\dot{\hat{\rho}} = -i
[\hat{H}_s,\hat{\rho}] - \frac{1}{2} \sum_{\mu=1}^2
\{\hat{L}_{\mu}^{\dag} \hat{L}_{\mu},\hat{\rho}\}+
\sum_{\mu=1}^2 \hat{L}_{\mu} \hat{\rho} \hat{L}_{\mu}^{\dag} \equiv \Lambda \rho.
\label{lindblad}
\end{equation}
Here $\hat{H}_s=\hat{n}^2/2+V(\hat{x},t)$ is the system
Hamiltonian, \{\,,\,\} is the anticommutator, and $\hat{L}_{\mu}$ are the Lindblad operators
given by \cite{Dittrich}
\begin{equation}
\begin{array}{l}
\hat{L}_1 = g \sum_n \sqrt{n+1} \; |n \rangle \, \langle n+1|,\\
\hat{L}_2 = g \sum_n \sqrt{n+1} \; |-n \rangle \, \langle -n-1|,
\end{array}
\end{equation}
with $n=0,1,...$ and $g=\sqrt{-\ln \gamma}$ (due to the Ehrenfest theorem).

The classical densities in phase space evolve with the
Perron-Frobenius operator arising from the Liouville equation
corresponding to the map (1). A coarse grained approximation to
the Perron-Frobenius is obtained by means of the Ulam method,
based on a discretization of the phase space. To construct the
Ulam matrix $S$, the phase space is divided into $M^2$ cells and
then $n_{\rm tr}$ random points from each cell $j$ are propagated
according to the classical map. The elements $S_{ij}$ of the $M^2
\times M^2 $ matrix $S$ are given by $ S_{ij} = {n_{ij} \over
n_{\rm tr}}$, where $n_{ij}$ is the number of trajectories
arriving to cell $i$ from the cell $j$. This discretization
introduces an effective diffusive noise of order $h_{\rm eff}^{\rm
PF} \propto {1 \over M}$ . For homogeneous systems and
sufficiently large values of $M$ the Ulam method is expected to
converge to the spectrum of the continuous system. When
thermal noise is included in the classical calculation we have
checked that the results obtained from the diagonalization
are independent of the value of $h_{\rm
eff}^{\rm PF}$ as long as $h_{\rm eff}^{\rm PF} \le h_{\rm eff}$,
i.e., the coarse graining inherent to the Ulam procedure is
smaller than the thermal fluctuations. In the following we will
refer to the Perron-Frobenius operator but the calculations are
understood to be done with its Ulam approximation.

In the quantum case the evolution of the density matrix is given by $\rho_{t+1} = e^ {\Lambda} \rho_{t}$,
where $e^ {\Lambda}$ is a non-unital superoperator of dimension $N^2 \times N^2$ constructed
by numerical integration of Eq. \ref{lindblad}. Here $h_{\rm eff} \propto {1 \over N}$.

For the diagonalization of $S$ (with and without thermal noise) and $e^
{\Lambda}$ we have used the Arnoldi method, which allowed us to go
to large dimensions, corresponding to  $0.15 < h_{\rm eff}^{\rm
PF} < 0.247$ for the classical matrix and $0.082 <h_{\rm eff}
<0.247$ for the quantum one. We should point out that these values
mean diagonalizing matrices of a maximum size of
$160\;000\;\times\;160\;000$ for the classical case (which takes
longer to construct) and of $531\;441\;\times\;531\;441$ for the
quantum case. Of course, we have compared similar $\hbar_{\rm
eff}^{\rm (PF)}$ sizes in both cases, expecting to improve the
construction time for the classical matrix in the future, in order
to compare with the already obtained quantum results.

\section{Classical to quantum correspondence from spectra}
\label{sec3}

\begin{figure}
\includegraphics[width=0.47\textwidth]{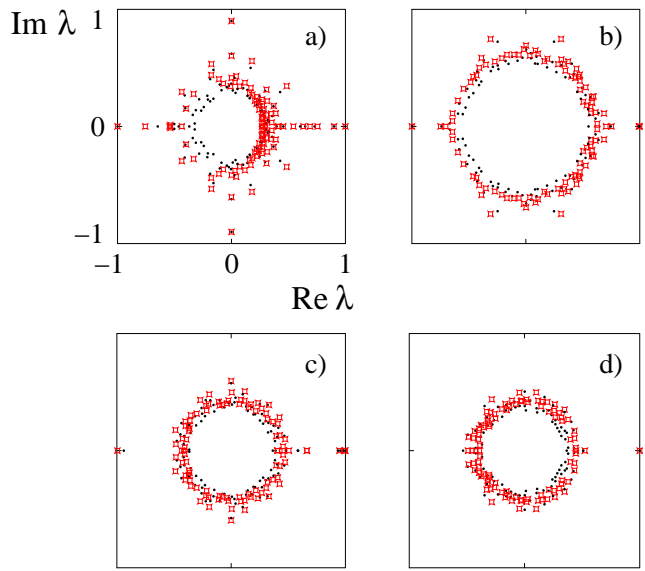}
\caption{(Color online) 100 largest eigenvalues of the
Perron-Frobenius operator $ S $ for the a) $B_{1}$, b) $C_{-1}$,
c) $D_{-1}$, and d) chaotic attractor cases. (Red) gray squares
correspond to $\hbar_{\rm eff}^{\rm PF}=0.15$, while black dots to
$\hbar_{\rm eff}^{\rm PF}=0.247$.} \label{fig1}
\end{figure}

We will first look at the properties of the spectra of the Perron-Frobenius operator
for the representative cases studied in \cite{Carlo}. In Fig. \ref{fig1} a) we show
the $B_{1}$ case ($k= 8.2$, $\gamma = 0.2$), in b) the $C_{-1}$ case ($k=5.6$, $\gamma = 0.64$),
in c) the $D_{-1}$ case ($k=11.9$, $\gamma =0.29$), and in d) the chaotic attractor case
($k=11.9$, $\gamma = 0.26$). We take $a=0.5$ and $\phi=\pi/2$ throughout this work.
By construction of $S$ these spectra are contained in the unit
circle and have a non-degenerate eigenvalue $\lambda_0=1$
\cite{Brin}. We compare the eigenvalues obtained for $\hbar_{\rm
eff}^{\rm PF}=0.15$ with those for $\hbar_{\rm eff}^{\rm
PF}=0.247$.

The first important thing we notice is that the eigenvalues that
are closer to the unit circle remain approximately fixed as
$\hbar_{\rm eff}^{\rm PF}$ decreases.
Moreover, we have checked that for $\hbar=0.15$ the right eigenvectors corresponding to
$\lambda_0$ reproduce the structure of the asymptotic classical
distributions (see Figs. 1 and 2 in \cite{Carlo}) in the cases a),
b) and c) which are point like ISSs, and the fractal chaotic
attractor in d). This strongly suggests that the actual classical dynamics
is already reasonably captured. It is interesting to notice that for the largest
$\hbar_{\rm eff}^{\rm PF}=0.247$,  this does not happen in the case $D_{-1}$,
where the expected point like structure is slightly embedded in
a distribution resembling the neighboring chaotic attractor case.
This indicates that in this region of the parameter space the greater
coarse graining is still not sufficient to resolve neighboring
maps with different properties.

For the ISSs cases, $\lambda_1$ (the eigenvalue following
$\lambda_0$ in decreasing order of their moduli) is very close to
$1$ in accordance with the very long times required for the
equilibration of these structures. As can be seen from Table
\ref{TableEquilibrationTimes} the decay times
$t_{\lambda_1}$ given essentially by the value of the spectral gap
$1-|\lambda_1|$, are in fairly good agreement with the
equilibration times obtained in \cite{Carlo} by means of the
ratchet currents.
\begin{table}[ht]
\begin{tabular}{l|cc|cc|cc|}
\cline{2-7}
       &\multicolumn{2}{|c|}{$PF$} &\multicolumn{2}{|c|}{$PF_{\rm th}$}  &\multicolumn{2}{|c|}{$QM$} \\
\cline{2-7}
       & $\lambda_1$ & $t_{\lambda_1}$  & $\lambda_1$ & $t_{
       \lambda_1}$        & $\lambda_1$ & $t_{\lambda_1}$  \\
\hline
\multicolumn{1}{|c|}{$B_{-1}$}    &   0.900       &   43.7      &   0.840      &   26.4            & 0.818        & 22.9        \\
\hline
\multicolumn{1}{|c|}{$C_{-1}$}    &   0.994       &  765.2      &   0.727      &   14.4            & 0.701        & 13.0        \\
\hline
\multicolumn{1}{|c|}{$D_{-1}$}    &   0.992       &  573.3      &   0.449      &   5.7             & 0.376        & 4.7         \\
\hline
\multicolumn{1}{|c|}{attr}      &   0.523       &   7.1       &   0.452      &   5.8             & 0.410        & 5.2        \\
\hline
\end{tabular}
\caption{Perron-Frobenius $\lambda_1$ values and equilibration
times $t_{\lambda_1}$ (defined by $\lambda_1^{t_{\lambda_1}}=0.01
$) are shown in columns 1 and 2 for all 4 cases analyzed in the
main text. Columns 3 and 4 show the same for the Perron-Frobenius
with thermal noise, and columns 5 and 6 for the quantum mechanical
case. We take $\hbar_{\rm eff}^{\rm (PF)}=\hbar_{\rm eff}= 0.15$.}
\label{TableEquilibrationTimes}
\end{table}
Again, we notice that in the $D_{-1}$ case the time $t_{\lambda_1}$
characterizing the decay towards the complex invariant state,
although large, is substantially shorter than the equilibration
time of about $700$ steps obtained with the exact classical dynamics \cite{Carlo}.

\begin{figure}
\includegraphics[width=0.47\textwidth]{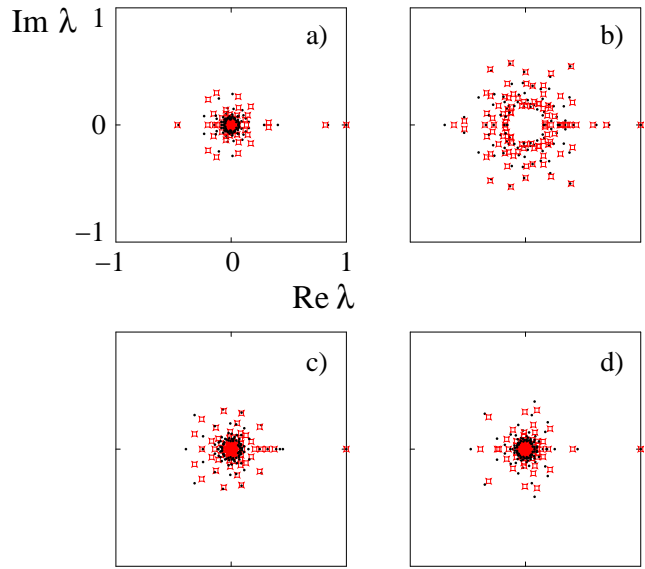}
\caption{(Color online) 100 largest eigenvalues for the
quantum superoperator $e^ {\Lambda}$ and for the Perron-Frobenius
operator with thermal noise $ S^{\rm thermal} $ for the a)
$B_{1}$, b) $C_{-1}$, c) $D_{-1}$, and d) chaotic attractor cases.
(Red) gray squares correspond to the quantum model, while black
dots to the classical one. $\hbar_{\rm eff}^{\rm (PF)}=\hbar_{\rm
eff}=0.15$.} \label{fig2}
\end{figure}

On the other hand, as can be checked with the help of Fig.
\ref{fig2} which displays the corresponding quantum spectra, the
bare coarse graining (even the greatest one $\hbar_{\rm eff}^{\rm
PF}=0.247$) involved in the Ulam procedure is not enough to
reproduce the quantum features in any of the studied ISSs, not
even in the $D_{-1}$ nor in the attractor case. In this Figure, the
quantum spectra for $\hbar_{\rm eff}=0.15$ ((red) gray squares)
are shown together with the classical results obtained from the
diagonalization of the Peron Frobenius operator with a thermal
noise satisfying $<\xi^2> = \hbar_{\rm eff}$ (black dots).
It is clear that with the help of the temperature the classical
spectra dramatically change in all cases and become remarkably
similar to those of the quantum system for the set of eigenvalues
of modulus closer to $1$.  Even so, it should be noticed that the
quantum dynamics is more contractive in spectral terms than the
classical one. In fact, a huge fraction of the quantum eigenvalues
have absolute values very close to 0. This was already pointed out
in \cite{baker} for the dissipative baker map, where the
differences between a full quantization and a discretization
procedure were studied.

In all cases the spectral gap ``jumps'' to the ``correct'' quantum
mechanical one. This is a highly non-trivial effect that
underlines the classical and quantum similarities. The chaotic
attractor case presents a singular property, this being that the
value of the quantum mechanical $\lambda_1$ is in good agreement
with its classical counterparts, with or without thermal noise.
Therefore the $t_{\lambda_1}$ is about the same in all cases (see
last row of Table \ref{TableEquilibrationTimes}) confirming  the
results of Fig. 4 of \cite{Carlo} which shows that in the case of
the chaotic attractor, classical and quantum currents have similar
equilibration times. For the remaining eigenvalues the
correspondence is not exact, but the similarities are evident.

We observe that the spectra of $D_{-1}$ and the chaotic attractor
cases are very close. This deepens on the hypothesis that chaotic
attractors could be approximated by neighboring ISSs, suggesting
that the approximation could be extended to the superoperator
itself. As a matter of fact, from these results we can conjecture
that the correspondence is at the (super) operator level and that
there should exist an approximate formal equivalence between them,
at least in their long-lived sector, i.e. the block associated to
the largest eigenvalues. In the following we will identify some
limits to this conjecture.

\section{The phase space picture}
\label{sec4}

In order to give a complementary point of view to analyze this
classical to quantum correspondence mechanism we study the
phase space behavior. We will compare the right
eigenvectors of the Perron-Frobenius operator with thermal noise, with the
Weyl-Wigner symbol for the right eigenvectors of the
quantum superoperator. Regarding the former it is worth noting
that the invariant eigenstates (with $\lambda=1$) are real and
non negative, in agreement with the Perron-Frobenius theorem, and
can be interpreted as probability distributions in phase space. The
remaining eigenvectors are real (although non positive) for
eigenvalues on the real axis and complex otherwise.
In the quantum case the invariant eigenstates
have the property of being density matrices with $Tr(\hat{R}_{\lambda_0})=1$,
while the remaining ones are traceless.

The Weyl-Wigner representation improves on previous results
obtained by using Husimi distributions \cite{Carlo}, since we are
able to appreciate interference fringes (coherences) and finer
details of the quantum eigenvectors. Weyl-Wigner symbols for a $N$
dimensional Hilbert space are defined in a redundant $2N\times2N$
discrete phase space \cite{opetor}. This is formed by the grid of
points $x=\frac{1}{N}(a,b)$ with $a$ and $b$ semi integer numbers
running from $0$ to $N-1/2$. In this way, the Weyl-Wigner symbol
$R(x)$ of the operator $\hat{R}$ is obtained from its
 matrix elements in the coordinate representation as
\[
R(x)=\sum_{n=0}^{N-1}<q_{2b-n}|\hat{R}|q_{n}>\exp\left(\frac{i2\pi}{N}2a(b-n)\right).
\]
In order to get rid of redundancies and ``ghost images'' derived essentially from
the cylindrical topology of our phase space, we use a method that
has been developed by Arg\"uelles and Dittrich \cite{Ditt} consisting of
Fourier transforming the Weyl-Wigner symbol to its symplectic analogue,
known as the ``chord symbol''. Then, after performing a cut off for the
longer chords and antifourier transforming, the new Weyl-Wigner symbol with the
desired properties is obtained. This latter is supported by a $N\times N$ discrete phase space formed by
points $x=\frac{1}{N}(i,j)$ where now $i$ and $j$ are integer numbers
running from $0$ to $N-1$. This is the symbol we use in the following.

It is important noticing that the eigenvectors are defined within a global phase factor.
For the invariant eigenstates the
property of being a density matrix ensures that the Weyl-Wigner symbol is real.
However, differently from the classical case, the quantum symbols display
negative values due to interference fringes.
For the remaining traceless eigenvectors, adjusting the phase factor,
we obtained that, in accordance with the classical case, the Weyl-Wigner
symbols are real for eigenvalues on the real axis and complex otherwise.
The complex eigenvalues come in complex conjugated pairs.
Their corresponding eigenvectors form pairs of complex conjugated
distributions in phase space also, both for the classical and the quantum case.

In Table \ref{TableOverlaps} we compare the overlaps between the
Weyl-Wigner symbol of the eigenvectors of the quantum
superoperator with the corresponding eigenvectors of the
Perron-Frobenius operator with thermal noise. We do this up to the
5th pair in decreasing order of the moduli of their corresponding
eigenvalues. To calculate these overlaps we take into account that
any state $\hat{R}$ can be represented by $R(x)$ with $x=(p,q)$ a
point in phase space. For the classical states, $R(x)$ stands for
the right eigenvector, while for the quantum ones, $R(x)$ is the
Weyl-Wigner symbol. Hence, given any two states $\hat{R_{1}}$ and
$\hat{R_{2}}$ , their overlap is defined as:
\[
O(\hat{R_{1}},\hat{R_{2}})=Tr\left(\hat{R_{1}}\hat{R_{2}}\right)/\sqrt{
\left[Tr\left(\hat{R_{1}^{2}}\right)Tr\left(\hat{R_{2}^{2}}\right)\right]}=
\]
\[
\sum_{x}R_{1}(x)R^{*}_{2}(x)/\sqrt{\left[\left(\sum_{x}|R_{1}(x)|^{2}\right)
\left(\sum_{x}|R_{2}(x)|^{2}\right)\right]},
\]
where $R^{*}(x)$ and $|R(x)|$ stand respectively for the complex
conjugate and absolute value of $R(x)$. The overlap defined above
is a complex magnitude, its modulus is invariant even though its
argument depends on the relative phase between the eigenvectors.
Also, when this relative phase is null
$O(\hat{R_{1}},\hat{R_{2}})$ is real. In all cases the value of
the overlap is about 0.9 for the invariant states and
progressively decreases as we go to smaller $|\lambda|$.
\begin{table}[ht]
\begin{centering}
\par\end{centering}
\begin{tabular}{|l|l|c|c|c|}
\hline
 & $B_{-1}$ \,  & \, $C_{1}$ \,  & \, $D_{1}$ \,  & Attr\tabularnewline
\hline
\hline
\, $\lambda_{0}$ \,  & $0.9449$ & $0.9349$ & $0.8697$ & $0.8654$ \tabularnewline
\, $\lambda_{1}$ \,  & $0.9441$ & $0.8132$ & $0.5422$ & $0.5186$ \tabularnewline
\, $\lambda_{2}$ \,  & $0.5622$ & $0.8504$ & $0.6381$ & $0.6689$\tabularnewline
\, $\lambda_{3}$ \,  & $0.5681$ & $0.8505$ & $0.6382$ & $0.6689$\tabularnewline
\, $\lambda_{4}$ \,  & $0.4421$ & $0.8141$ & $0.5190$ & $0.4178$\tabularnewline
\hline
\end{tabular}
\caption{Overlaps of the first 5 eigenvectors of the quantum
superoperator with the corresponding eigenvectors of the
Perron-Frobenius operator with thermal noise, for all cases.}
\label{TableOverlaps}
\end{table}

In the following Figures we display the phase space portraits of selected eigenvectors
of the Perron-Frobenius operator with thermal noise and the Weyl-Wigner symbols of the
corresponding eigenvector of the quantum superoperator.

\begin{figure}
\includegraphics[width=0.47\textwidth]{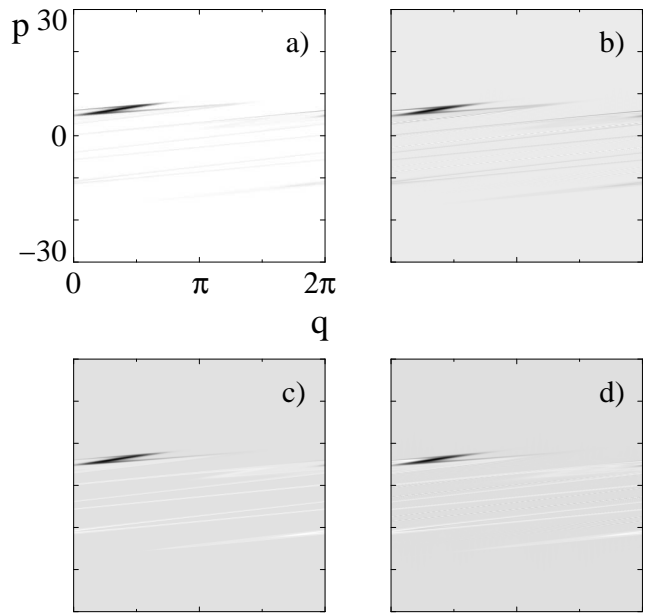}
\caption{Phase space portraits of two eigenvectors of the
Perron-Frobenius with thermal noise (left panels) and the
Weyl-Wigner symbols for the corresponding quantum eigenstates
(right panels), for the $B_{1}$ ISS. In a) and b) we show the
eigenvectors associated with $\lambda_0$, while in c) and d) the
ones corresponding to $\lambda_1$ ($\lambda_1=0.840$ in the
classical spectrum and $\lambda_1=0.818 $ in the quantum one). In
all cases $\hbar_{\rm eff}^{\rm (PF)}=0.15$.} \label{fig3}
\end{figure}

We first analyze the $B_{1}$ ISS. This is the only case where the
invariant distributions shown in Fig.\ref{fig3} a) and b)
correspond to point-like attractors, even though some traces of
the chaotic basin of attractor are visible in both panels. As can
be seen from Fig.\ref{fig3} a) a smearing out of the order of
$h_{\rm eff}$  is enough to turn the classical points
corresponding to the ISS into a distribution of exactly the same
shape of the quantum one depicted in Fig.\ref{fig3} b).  The
quantum-classical similarity is also striking for the 2nd
eigenvector pair displayed in Fig.\ref{fig3} c) and d) . Despite
both having a negative image of the basin of attraction these
states behave in a very similar way to the invariant pair.

\begin{figure}
\includegraphics[width=0.47\textwidth]{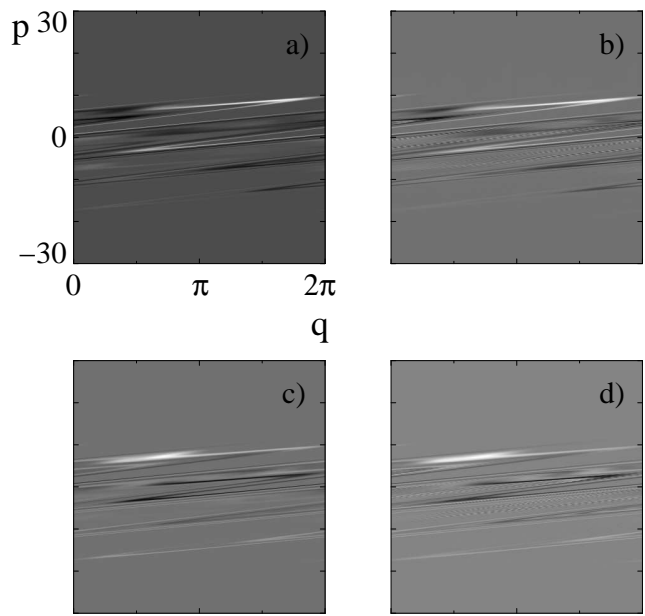}
\caption{Phase space portrait of the eigenvector associated with
the complex eigenvalue $\lambda_2$ ($\lambda_2 = 0.388 + {\rm i}
0.500 $ in the classical spectrum and $\lambda_2=0.396+ {\rm i}
0.501$ in the quantum one) of the Perron Frobenius with thermal
noise (left panels) and the Weyl-Wigner symbols for the
corresponding quantum eigenstates (right panels) for the $C_{-1}$
ISS. In a) and b) we show the real part of the distributions while
in c) and d) the imaginary one. In all cases $\hbar_{\rm eff}^{\rm
(PF)}=0.15$.} \label{fig4}
\end{figure}

For the next case corresponding to the $C_{-1}$ ISS, we have
chosen to show the details of the eigenvectors associated with
complex $\lambda_2$.  The real parts shown in Fig. \ref{fig4} a)
and b) are different from the imaginary ones shown in Fig.
\ref{fig4} c) and d), and again the coincidence between classical
and quantum phase space portraits is remarkable. This is also
clearly reflected in the value of the overlap.  The main
difference is in the presence of interference fringes in the
quantum case which do not have enough weight to spoil the overlap
in a significant way.

\begin{figure}
\includegraphics[width=0.47\textwidth]{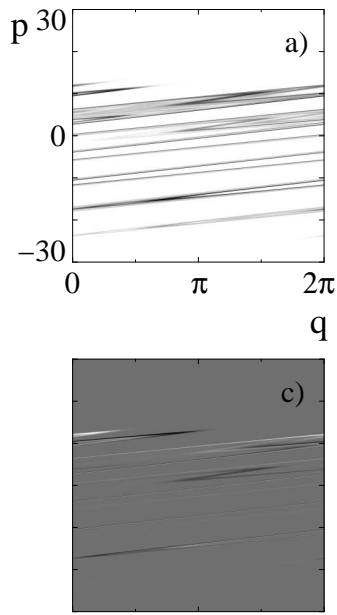}
\caption{Phase space portraits of two eigenvectors of the
Perron-Frobenius with thermal noise (left panels) and the
Weyl-Wigner symbols for the corresponding quantum eigenstates
(right panels), for the $D_{-1}$ ISS. In a) and b) we show the
eigenvectors associated with $\lambda_0$, while in c) and d) the
ones corresponding to $\lambda_4$ ($\lambda_4=0.391$ in the
classical spectrum and $\lambda_4=0.313 $ in the quantum one). In
all cases $\hbar_{\rm eff}^{\rm (PF)}=0.15$.} \label{fig5}
\end{figure}

\begin{figure}
\includegraphics[width=0.47\textwidth]{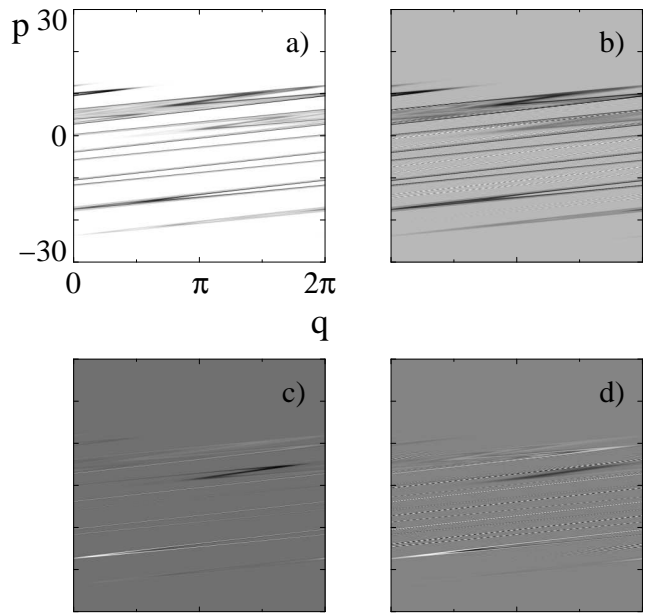}
\caption{Phase space portraits of two eigenvectors of the
Perron-Frobenius with thermal noise (left panels) and the
Weyl-Wigner symbols for the corresponding quantum eigenstates
(right panels), for the chaotic attractor. In a) and b) we show
the eigenvectors associated with $\lambda_0$, while in c) and d)
the ones corresponding to $\lambda_4$ ($\lambda_4=-0.471$ in the
classical spectrum and $\lambda_4=-0.391 $ in the quantum one). In
all cases $\hbar_{\rm eff}^{\rm (PF)}=0.15$.} \label{fig6}
\end{figure}

The next two cases are better analyzed together since they belong to the same region of the
parameter space. It can be seen that they coincide, having very similar invariant
states shown in Figs. \ref{fig5} and \ref{fig6}, panels a) and b).
The differences are only present through interference fringes at the quantum level.
But they are clearly not very relevant, both looking at these representations and
also at the values of the overlaps.
On the other hand, we show the eigenvectors corresponding to the worst performing overlaps
in both cases in Figs. \ref{fig5} and \ref{fig6}, panels c) and d).
They show that in these cases the coherences have a greater weight. Anyway if we look carefully
at the distributions they are quite alike for the remaining of the distributions.
This puts a limit on the previously mentioned hypothesis consisting of approximating the
long lived block of the quantum superoperators with the Perron-Frobenius ones with thermal noise.
A quantitative study is left for the future.

\section{Conclusions}
\label{sec5}

By studying the behavior of the spectra and eigenvectors of both
the Perron-Frobenius operators with thermal noise and the quantum mechanical
superoperators associated with the ISSs and chaotic attractors
typical of a paradigmatic model for dissipative directed
transport, we were able to identify a novel mechanism of classical
to quantum correspondence hinted in \cite{Carlo} (just looking at
the ratchet currents and asymptotic distributions).
The remarkable coincidence of both sets of eigenvalues is
indicative of an approximate formal equivalence between the
classical and quantum realms. In the following we raise
some important points derived from this result.

Firstly, it is interesting to link our approach with previous
works. In \cite{Braun} the dissipative kicked top model was
analyzed, with a special focus in the case where it decays to a
chaotic attractor. There it was found that the quantum invariant
state closely follows the structure of the classical attractor,
without the finer details. It was also shown that in the
semiclassical limit the leading eigenvalues of the quantum
spectrum converge to the ones of the Perron-Frobenius, thus
implying that the quantum mechanical time scales become
independent of $h_{\rm eff}$ and coincide with time scales set by
the Ruelle resonances. In the present work the classical to
quantum correspondence is investigated in representative regions
of the parameter space of a paradigmatic dissipative ratchet
system, corresponding to chaotic attractors and to ISSs as well.
Besides, we put the accent on small but finite $\hbar_{\rm eff}$
values, for which we find a correspondence mechanism based on
adding thermal noise at the classical level only. Our approach is
also clearly distinct from the usual way in which the quantum to
classical correspondence is treated in the decoherence literature
where the procedure consists in applying a diffusive noise to both
the classical and the quantum equations in order for the spectra
to converge to each other \cite{nonnen}.

On the other hand, the consequences of this finding are relevant
not only to directed transport but to dissipative systems in
general. A very promising one is the possibility to approximate the quantum
asymptotic state in the chaotic case by the classical dynamics with thermal
noise corresponding to a simple attractor in its vicinity. This was
previously conjectured in \cite{Carlo} but now with the help of the Weyl-Wigner
distributions we are able to show that this is indeed possible. As a matter
of fact, the differences due to coherences are minor.
Moreover, it is worth noticing that even when having a classical chaotic
attractor the addition of thermal noise at this level is necessary in
order to find a spectrum similar to the quantum one. Hence,
studying these effects could be interesting for many applications
as for example in superconducting quantum interference devices (SQUIDs)
\cite{Everitt}.

Finally, we would like to point out that the long-lived sector of the quantum
superoperators corresponding to any kind of structure in parameter space
could be well approximated by
means of Perron-Frobenius operators with thermal noise.
This conclusion can be drawn from the remarkable similarity among the spectra for all
the cases shown in Sec. \ref{sec3}. However, in our way to devise a classical approximation
we have found a problem hardly visible using just Husimi distributions as in
\cite{Carlo}. Thanks to the Weyl-Wigner representations of Sec. \ref{sec4} we have found
that the eigenvectors, even some long-lived ones, can have differences due to
the coherences remaining in the quantum system.
Despite this, their weight seems to be not so important.
This keeps our hopes to find an effective approximation theory in the future.

\section*{Acknowledgments}

Support from CONICET is gratefully acknowledged.

\vspace{3pc}



\begin{thebibliography}{99}

\bibitem{Feynman}
R. P. Feynman,
{\it Lectures on Physics}, {\bf Vol. 1},
(Addison-Wesley,  Reading, MA, 1963).

\bibitem{Reimann}
P. Reimann,
Phys. Rep. {\bf 361}, 57 (2002).

\bibitem{Kohler}
S. Kohler, J. Lehmann, and P. H\"anggi, Phys. Rep. {\bf 406}, 379
(2005).

\bibitem{Denisov}
S. Denisov, S. Flach, and P. H\"anggi,
Phys. Rep. {\bf 538}, 77 (2014).

\bibitem{biology}
F. J\"ulicher, A. Ajdari and J. Prost,
Rev. Mod. Phys. {\bf 69}, 1269 (1997);
G. Mahmud \emph{et al.},
Nature Phys. {\bf 5}, 606 (2009);
G. Lambert, D. Liao, and R.H. Austin,
Phys. Rev. Lett. {\bf 104}, 168102 (2010).

\bibitem{nanodevices}
R. D. Astumian,
Science {\bf 276}, 917 (1997);
D. Reguera, A. Luque, P.S. Burada, G. Schmid, J.M. Rub\'\i ,and P. H\"anggi,
Phys. Rev. Lett. {\bf 108}, 020604 (2012).

\bibitem{chemistry}
J. B. Gong and P. Brumer,
Annu. Rev. Phys. Chem. {\bf 56}, 1 (2005);
G. G. Carlo, L. Ermann, F. Borondo, and R. M. Benito,
Phys. Rev. E {\bf 83}, 011103 (2011).

\bibitem{CAexp}
P. H. Jones, M. Goonasekera, D.R. Meacher, T. Jonckheere, and T.S. Monteiro,
Phys. Rev. Lett. {\bf 98}, 073002 (2007);
T. Salger, S. Kling, T. Hecking, C. Geckeler, L. Morales-Molina, and M. Weitz,
Science {\bf 326}, 1241 (2009).

\bibitem{AOKR}
T. S. Monteiro, P. A. Dando, N. A. C. Hutchings, and M. R. Isherwood,
Phys. Rev. Lett. {\bf 89}, 194102 (2002);
G. G. Carlo, G. Benenti, G. Casati, S. Wimberger, O. Morsch, R. Mannella, and E. Arimondo,
Phys. Rev. A {\bf 74}, 033617 (2006).

\bibitem{BECratchets}
M. Sadgrove, M. Horikoshi, T. Sekimura, and K. Nakagawa,
Phys. Rev. Lett. {\bf 99}, 043002 (2007);
I. Dana, V. Ramareddy, I. Talukdar, and G.S. Summy,
Phys. Rev. Lett. {\bf 100}, 024103 (2008).

\bibitem{purelyQR}
E. Lundh and M. Wallin,
Phys. Rev. Lett. {\bf 94}, 110603 (2005);
D. Poletti, G. G. Carlo, and B. Li,
Phys. Rev. E {\bf 75}, 011102 (2007).

\bibitem{QR2}
A. Kenfack, J. Gong, and A.K. Pattanayak,
Phys. Rev. Lett. {\bf 100}, 044104 (2008);
J. Wang and J. Gong,
Phys. Rev. E {\bf 78}, 036219 (2008).

\bibitem{coherentControl}
M. Sadgrove, M. Horikoshi, T. Sekimura, and K. Nakagawa,
Eur. Phys. J. D {\bf 45}, 229 (2007).

\bibitem{origin}
S. Flach, O. Yevtushenko, and Y. Zolotaryuk,
Phys. Rev. Lett. {\bf 84}, 2358 (2000).

\bibitem{Mateos}
J. L. Mateos,
Phys. Rev. Lett {\bf 84}, 258 (2000).

\bibitem{qdisratchets}
G. G. Carlo, G. Benenti, G. Casati, and D.L. Shepelyansky,
Phys. Rev. Lett. {\bf 94}, 164101 (2005).

\bibitem{Celestino}
A. Celestino, C. Manchein, H.A. Albuquerque, and M.W. Beims,
Phys. Rev. Lett. {\bf 106}, 234101 (2011).

\bibitem{Manchein}
C. Manchein, A. Celestino, and M.W. Beims,
Phys. Rev. Lett. {\bf 110}, 114102 (2013).

\bibitem{Robnik}
D.F.M. Oliveira, M. Robnik, and E.D. Leone,
Chaos {\bf 21}, 043122 (2011).

\bibitem{Carlo}
G.G. Carlo,
Phys. Rev. Lett. {\bf 108}, 210605 (2012).

\bibitem{Ermann}
L. Ermann and G.G. Carlo,
Phys. Rev. E {\bf 91}, 010903(R) (2015).

\bibitem{nonnen}
S. Nonnenmacher,
Nonlinearity {\bf 16}, 1685 (2003).

\bibitem{Zurek}
W.H. Zurek,
Rev. Mod. Phys. {\bf 75}, 715 (2003).

\bibitem{Lindblad}
G. Lindblad,
Commun. Math. Phys. {\bf 48}, 119 (1976).

\bibitem{Dittrich}
T. Dittrich and R. Graham,
Europhys. Lett., {\bf 7}, 287 (1988).

\bibitem{Brin}
Introduction to Dynamical Systems, M. Brin and G. Stuck, Cambridge
University Press, Cambridge UK, 2002.

\bibitem{baker}
G. G. Carlo, A. M. F. Rivas, and M. E. Spina,
Phys. Rev. E {\bf 84}, 066201 (2011).

\bibitem{opetor} A.M.F. Rivas and A.M. Ozorio de Almeida,
Annals of Physics {\bf 276}, 223, (1999).

\bibitem{Ditt} A. Arg\"uelles and T. Dittrich,
Physica A {\bf 356}, 72, (2005).

\bibitem{Braun}
Dissipative Quantum Chaos and Decoherence,
D. Braun,
Springer Tracts in Modern Physics 172,
Springer, Berlin Heidelberg New York, 2001.

\bibitem{Everitt}
M.J. Everitt,
New J. of Phys. {\bf 11}, 013014 (2009).

\end{thebibliography}
\end{document}